# Algebraic decomposition of the electromagnetic constitutive tensor.

## A step towards a pre-metric based gravitation?

E. Matagne, Université catholique de Louvain
3, pl. du Levant, B-1348 Louvain-la-Neuve
http://www.lei.ucl.ac.be/~matagne/matagne.html
ernest.matagne@uclouvain.be
++32/(0)10 47 22 61

## Abstract

The aim of this paper is to provide an algebraic decomposition of the principal part of the electromagnetic constitutive tensor $\chi_{\mu\nu\rho\sigma}$ (a pre-metric structure), in a sum of terms of the form $\sum_i Z_{(i)} \left( g_{\mu\rho}^{(i)} g_{\nu\sigma}^{(i)} - g_{\mu\sigma}^{(i)} g_{\nu\rho}^{(i)} \right)$ where the $Z_{(i)}$ are invariant scalars and $g_{\mu\nu}^{(i)}$ unimodular conformal metrics. If the space-time is also equipped with a volume measure, such a decomposition is equivalent to a similar decomposition of the associated area metric, where the $g_{\mu\nu}^{(i)}$ are ordinary metrics. It has been recently argued [8] that, if such a particular decomposition should be available, it would then be possible to find a compatible connection and derived tensors.

## Notations

We use in this paper the component language. The Greek indices are taking value in $\{0, 1, 2, 3\}$. We consider only frames with one orientation, and thus there is in this paper no distinction between tensors and pseudo-tensors. We make use of the Einstein's dumb indices convention.

## Introduction

Following the Maxwell-Minkowski-Post formalism [2], electromagnetics is based on two tensorial quantities, the Maxwell-Faraday tensor $F_{\mu\nu}$ and the Maxwell-Ampère tensor, here denoted by $H^{\mu\nu}$. $F_{\mu\nu}$ is a skew-symmetric tensor with two covariant indices, and $H^{\mu\nu}$ is a skew-symmetric tensorial density with two contravariant indices. Equivalently, $F_{\mu\nu}$ is a 2-form and $H^{\mu\nu}$ is the dual of another 2-form. Each of these quantities obeys to separate equations of evolution (respectively homogeneous and inhomogeneous Maxwell equations). They are linked only by a constitutive relation witch, in the more general linear case, is of the form

$$(1) \quad F_{\mu\nu} = \frac{1}{2} \chi_{\mu\nu\rho\sigma} H^{\rho\sigma} \quad .$$

The tensor $\chi_{\mu\nu\rho\sigma}$ used in (1) is the reciprocal of the tensor $\chi^{\mu\nu\rho\sigma}$ used by Post [2]. This doesn't cause trouble thanks to the electromagnetic self-duality. We use here the formulation (1) in order to have a tensor with the same index structure as the area metric tensor [8].

The electromagnetic constitutive tensor $\chi_{\mu\nu\rho\sigma}$ is a tensor of weight -1 , which present the symmetries

(2)  $\chi_{\mu\nu\rho\sigma} = -\chi_{\nu\mu\rho\sigma} = -\chi_{\mu\nu\sigma\rho}$  .

We limit ourselves here to a medium in which the energy is conserved, so that one admits also the symmetry

(3) $\chi_{\mu\nu\rho\sigma} = \chi_{\rho\sigma\mu\nu}$  .

In the general case, the tensor $\chi_{\mu\nu\rho\sigma}$ can be decomposed in three irreductible parts: the principal part, the axion part and the skewon part [5] [7]. Owing to the symmetry (3), we don't consider here the so-called "skewon part" of the constitutive tensor. The tensor here studied is thus the sum of the principal part and of the axion part. It has 21 independent components. The interpretation of the above quantities after time-space splitting is given in Appendix 1.

In general relativity, one has in the vacuum

(4)   $\chi_{\mu\nu\rho\sigma} = Z_o \dfrac{1}{\sqrt{g}} (g_{\mu\rho} \, g_{\nu\sigma} - g_{\mu\sigma} \, g_{\nu\rho})$   ,

where $g_{\mu\nu}$ is the metric, g the absolute value of its determinant and $Z_o$ a universal constant (the impedance of free space $\approx 377\ \Omega$).

Thus, the tensor $\chi_{\mu\nu\rho\sigma}$ is a natural generalisation of the metric, and carries more degrees of freedom (21 instead of 10). For that reason, one says sometime that it defines a "pre-metric" structure [5] [7]. It's possibly a way to find a geometrical theory of more physical phenomena as gravitation.

If the space-time is equipped with a volume measure $\eta$ (a density, i.e. a scalar of weight 1), then  $\eta\ \chi_{\mu\nu\rho\sigma}$ is an ordinary tensor (tensor of weight 0 ), which still presents the symmetries (2)(3). Such a tensor has been called " algebraic curvature tensor" [4] or "area metric" [8].

Gilkey [4] has proved that, except for the totally skew-symmetric part, such a tensor can be split in a sum of terms following

(5)        $\eta\ \chi_{\mu\nu\rho\sigma} = \sum_{i} \underset{(i)}{\sigma}(\underset{(i)}{g_{\mu\rho}}\ \underset{(i)}{g_{\nu\sigma}} - \underset{(i)}{g_{\mu\sigma}}\ \underset{(i)}{g_{\nu\rho}})$   ,

where the  $\underset{(i)}{g_{\mu\nu}}$  are symmetric tensors with 2 covariant indices (metrics).
The aim of this paper is to provide a way to obtain that decomposition. It has indeed been recently showed [8] that, if such a decomposition were known, it should be possible to construct a compatible connection and derived tensors.

## Use of the permutation tensor

The permutation tensor $\varepsilon_{\mu\nu\rho\sigma}$ is a completely anti-symmetric tensor with four covariant indices. It has the same components in all frames and is thus trivially defined by

(6)        $\varepsilon_{0123} = 1$  .

That tensor $\varepsilon_{\mu\nu\rho\sigma}$ has the same weight as the constitutive tensor $\chi_{\mu\nu\rho\sigma}$, and obeys also the symmetry (2)(3).

As the tensor (6) exists on all manifolds, it can be used anyway without introducing additional structure in the problem treated. One says that the tensor is "trivial".

## A reminder of 6-vectors

The quantities $F_{\mu\nu}$ and $H^{\mu\nu}$ have only six independent components, as one can see in Appendix I.

Then, if we note by upper-case Latin letter indices with values in {01, 02, 03, 23, 31, 12}, we can replace these quantities by $F_A$ and $H^A$. $F_A$ and $H^A$ are vectors of a six dimensional "vector space". They are called 6-vectors.

The tensors $\chi_{\mu\nu\rho\sigma}$ and $\varepsilon_{\mu\nu\rho\sigma}$ are replaced by $\chi_{AB}$ and $\varepsilon_{AB}$. The components of these tensors form then 6 x 6 matrices. By example, we have

$$(7) \qquad \varepsilon_{AB} = \begin{bmatrix} 0 & 0 & 0 & 1 & 0 & 0 \\ 0 & 0 & 0 & 0 & 1 & 0 \\ 0 & 0 & 0 & 0 & 0 & 1 \\ 1 & 0 & 0 & 0 & 0 & 0 \\ 0 & 1 & 0 & 0 & 0 & 0 \\ 0 & 0 & 1 & 0 & 0 & 0 \end{bmatrix} \ .$$

We recall in Appendix II the form of $\chi_{AB}$ for vacuum in ordinary co-ordinates (t, x, y, z).

The space of 6-vectors has invariance properties. For each change of frame of matrix $A^{\mu}{}_{\mu'}$, we have indeed a transformation

$$(8) \qquad T^A{}_{A'} = A^{[\mu}{}_{[\mu'} A^{\nu]}{}_{\nu']} \ .$$

For any matrix $T^A{}_{A'}$ of the form (8), $\varepsilon_{AB}$ is invariant under the transformation

$$(9) \qquad \varepsilon_{A'B'} = T^{-1/3} \varepsilon_{AB} \, T^A{}_{A'} \, T^B{}_{B'} \ ,$$

where T is the determinant of $T^A{}_{A'}$.

Conversely, if a matrix $T^A{}_{A'}$ leaves $\varepsilon_{AB}$ invariant under (9), it corresponds by (8) to two transformations $A^{\mu}{}_{\mu'}$ which differ only by a factor -1.

The notion of 6-vector is useful for the algebraic manipulations on electromagnetic quantities. It is related to the notion of semi-vector introduced by Einstein and Mayer [1].

## Eigenvalues and eigenvectors of the constitutive tensor

The matrix (7) is symmetric. The tensor $\varepsilon_{AB}$ can thus play in the vector space of 6-vectors a role similar to the metric in the ordinary space. The main idea of this paper is to use that tensor $\varepsilon_{\mu\nu\rho\sigma}$ in order to analyse the constitutive tensor $\chi_{\mu\nu\rho\sigma}$ in terms of "eigenvectors" and "eigenvalues". By this way, as $\varepsilon_{\mu\nu\rho\sigma}$ is a trivial tensor, the **parsimony principle** will be fully satisfied.

Consider the equation in $\lambda$

(10)     $0 = \det\left(\chi_{AB} - \lambda\,\varepsilon_{AB}\right)$ .

The right member of that equation is a polynomial de degree 6. Thus, there are 6 solutions, which can be real or complex by conjugated pairs.

On the other hand, for each root $\lambda_{(i)}$ of (10), there exists an "eigenvector" $H^A_{(i)}$ such that

(11)     $\chi_{AB}\,H^B_{(i)} = \lambda_{(i)}\,\varepsilon_{AB}\,H^B_{(i)}$   .

It is easy to proof that, if $\lambda_{(i)} \neq \lambda_{(j)}$ , one has

(12)     $H^A_{(i)}\,\varepsilon_{AB}\,H^B_{(j)} = 0$   .

Assuming all the eigenvalues distinct and $H^A_{(i)}\,\varepsilon_{AB}\,H^B_{(i)} \neq 0$ , there is thus 6 independent eigenvectors.

We assume in the following that the roots of (10) are formed of three pairs of conjugated roots. We shall note $\lambda_1$ , $\lambda_2$ and $\lambda_3$ the three roots with positive imaginary part. The six roots of (10) are thus $\{\lambda_1, \lambda_2, \lambda_3, \lambda_1^*, \lambda_2^*, \lambda_3^*\} = \{\lambda_{(i)}, \lambda_{(i)}^*\}$ with $i \in \{1, 2, 3\}$. The real and imaginary parts of those roots are invariant scalars.

The eigenvectors can be normalised by the additional condition

(13)     $H^A_{(i)}\,\varepsilon_{AB}\,H^B_{(i)} = j2$   ,

where  $j = \sqrt{-1}$ .

Then, it is possible to decompose the constitutive tensor

(14)     $\chi_{AB} = \lambda_{(1)}\,H^C_{(1)}\,\varepsilon_{AC}\,\dfrac{1}{j\,2}\,\varepsilon_{BD}\,H^D_{(1)} + \lambda_{(2)}\,H^C_{(2)}\,\varepsilon_{AC}\,\dfrac{1}{j\,2}\,\varepsilon_{BD}\,H^D_{(2)} + \lambda_{(3)}\,H^C_{(3)}\,\varepsilon_{AC}\,\dfrac{1}{j\,2}\,\varepsilon_{BD}\,H^D_{(3)}$   .

+ complex   conjugate

## Return to real quantities

We denote

(15)     $H^A_{(i)} = H^A_{(i)r} + j\,H^A_{(i)i}$   with $i \in \{1, 2, 3\}$

an eigenvector associated with $\lambda_{(i)}$ . Then,

(16)     $H^{*A}_{(i)} = H^A_{(i)r} - j\,H^A_{(i)i}$

is an eigenvector associated with $\lambda_{(i)}^*$ . Owing to (12)(13), we obtain,

(17)     $H^A_{(i)r}\,\varepsilon_{AB}\,H^B_{(i)r} - H^A_{(i)i}\,\varepsilon_{AB}\,H^B_{(i)i} = 0$ ,

(18)     $H^A_{(i)r}\,\varepsilon_{AB}\,H^B_{(i)i} + H^A_{(i)i}\,\varepsilon_{AB}\,H^B_{(i)r} = 2$

and, if $\lambda_{(i)} \neq \lambda_{(k)}$ ,

$$(19) \quad H^A_{(i)r} \, \varepsilon_{AB} \, H^B_{(k)r} - H^A_{(i)i} \, \varepsilon_{AB} \, H^B_{(k)i} = 0 \quad ,$$

$$(20) \quad H^A_{(i)r} \, \varepsilon_{AB} \, H^B_{(k)i} + H^A_{(i)i} \, \varepsilon_{AB} \, H^B_{(k)i} = 0 \quad .$$

On the other hand, since $\lambda_{(i)}$ and $\lambda_{(k)}{}^*$ are distinct eigenvalues, one has for all values of i and k

$$(21) \quad H^A_{(i)r} \, \varepsilon_{AB} \, H^B_{(k)r} + H^A_{(i)i} \, \varepsilon_{AB} \, H^B_{(k)i} = 0 \quad ,$$

$$(22) \quad H^A_{(i)r} \, \varepsilon_{AB} \, H^B_{(k)i} - H^A_{(i)i} \, \varepsilon_{AB} \, H^B_{(k)r} = 0 \quad .$$

The conditions (17) to (22) are equivalent to

$$(23) \quad H^A_{(i)r} \, \varepsilon_{AB} \, H^B_{(i)i} = H^A_{(i)i} \, \varepsilon_{AB} \, H^B_{(i)r} = \alpha$$

and, for $i \neq k$,

$$(24) \quad H^A_{(i)r} \, \varepsilon_{AB} \, H^B_{(k)i} = H^A_{(i)i} \, \varepsilon_{AB} \, H_{(k)r} = 0$$

and, for all values of i and k,

$$(25) \quad H^A_{(i)r} \, \varepsilon_{AB} \, H^B_{(k)r} = H^A_{(i)i} \, \varepsilon_{AB} \, H^B_{(k)i} = 0 \quad .$$

Then, if we define

$$(26) \quad \lambda_{(i)} = \mu_{(i)} + j \, \nu_{(i)} \quad \text{with} \quad \nu_{(i)} > 0 \quad ,$$

the decomposition (14) can be written as a sum or real terms

$$(27) \quad \begin{aligned} \chi_{AB} = & \sum_i \mu_{(i)} \, ( \, H^C_{(i)r} \, \varepsilon_{AC} \, \varepsilon_{BD} \, H^D_{(i)i} \, + H^C_{(i)i} \, \varepsilon_{AC} \, \varepsilon_{BD} \, H^D_{(i)r} ) \\ & - \sum_i \nu_{(i)} \, ( \, H^C_{(i)r} \, \varepsilon_{AC} \, \varepsilon_{BD} \, H^D_{(i)r} - H^C_{(i)i} \, \varepsilon_{AC} \, \varepsilon_{BD} \, H^D_{(i)i} ) \end{aligned}$$

## Construction of the proper frame

If we build a matrix T the columns of which are formed of the components of the eigenvalues, namely

$$(28) \quad T^A{}_{A'} = \left[ \, H^A_{(1)r} \quad H^A_{(2)r} \quad H^A_{(3)r} \quad H^A_{(1)i} \quad H^A_{(2)i} \quad H^A_{(3)i} \, \right] \quad ,$$

then, it is obvious by (23) to (25) that

(29)     $\varepsilon_{A'B'} = \varepsilon_{AB} \, T^A_{\ A'} \, T^B_{\ B'}$   .

Taking the determinant of the two members of (31), we are led to the conclusion that $T^A_{\ A'}$ is unimodular

(30)   T = 1.

Thus, the condition (9) is fulfilled and there exist a frame transformation $A^\mu_{\ \mu'}$ corresponding to $T^A_{\ A'}$ .

On the other hand, using (23) to (25) and (27), one finds

(31)     $\chi_{AB} \, H^B_{(i)r} = \mu_{(i)} \, H^C_{(i)r} \, \varepsilon_{AC} - \nu_{(i)} \, H^C_{(i)i} \, \varepsilon_{AC}$   ,

(32)     $\chi_{AB} \, H^B_{(i)i} = \nu_{(i)} \, H^C_{(i)r} \, \varepsilon_{AC} + \mu_{(i)} \, H^C_{(i)i} \, \varepsilon_{AC}$   .

Using again (23) to (25), we obtain owing to (28) and (30)

(33)     $\chi_{A'B'} = T^{-1/3} \, T^A_{\ A'} \, \chi_{AB} \, T^B_{\ B'} = \begin{bmatrix} -\nu_{(1)} & 0 & 0 & \mu_{(1)} & 0 & 0 \\ 0 & -\nu_{(2)} & 0 & 0 & \mu_{(2)} & 0 \\ 0 & 0 & -\nu_{(3)} & 0 & 0 & \mu_{(3)} \\ \mu_{(1)} & 0 & 0 & \nu_{(1)} & 0 & 0 \\ 0 & \mu_{(2)} & 0 & 0 & \nu_{(2)} & 0 \\ 0 & 0 & \mu_{(3)} & 0 & 0 & \nu_{(3)} \end{bmatrix}$   .

The frame in which the constitutive tensor has the form (33) can be called the proper frame of the medium. The form (33) is insensible to a dilatation of that frame, which is normal because electromagnetism is conformally invariant.

Now, by analogy with (5), we will search for a decomposition

(34)     $\chi_{\mu\nu\rho\sigma} = \sum_i Z_{(i)} \, (g_{\mu\rho} \, g_{\nu\sigma} - g_{\mu\sigma} \, g_{\nu\rho})$   ,

where the $g_{\mu\nu}$ are conformal metrics, i.e. symmetric tensors of weight -1/2 , with two covariant indices. These tensors will be assumed to have a determinant equal to 1 or -1. That property is verified in all frames simultaneously.

The algebraic decomposition (34) of the constitutive tensor is easy when it is in the form (33), as we will see in the following. We are thus searching a decomposition

(35)     $\chi_{AB} = \sum_i Z_{(i)} \, G_{AB}_{(i)}$   ,

where

(36)     $G_{\mu\nu\rho\sigma}_{(i)} = g_{\mu\rho}_{(i)} \, g_{\nu\sigma}_{(i)} - g_{\mu\sigma}_{(i)} \, g_{\nu\rho}_{(i)}$   .

## Decomposition of the main diagonal of the constitutive tensor

Considering a constitutive tensor (33) with only the main diagonal, it can be decomposed in four terms of the form (35)(36), with

$$\text{(37a)} \quad g_{\mu\nu} \atop (0) = \text{diag}(-1, 1, 1, 1) \quad,$$

$$\text{(37b)} \quad g_{\mu\nu} \atop (1) = \text{diag}(1, -1, 1, 1) \quad,$$

$$\text{(37c)} \quad g_{\mu\nu} \atop (2) = \text{diag}(1, 1, -1, 1) \quad,$$

$$\text{(37d)} \quad g_{\mu\nu} \atop (3) = \text{diag}(1, 1, 1, -1)$$

and, thus,

$$\text{(38a)} \quad G_{AB} \atop (0) = \text{diag}(-1, -1, -1, 1, 1, 1) \quad,$$

$$\text{(38b)} \quad G_{AB} \atop (1) = \text{diag}(-1, 1, 1, 1, -1, -1) \quad,$$

$$\text{(38c)} \quad G_{AB} \atop (2) = \text{diag}(1, -1, 1, -1, 1, -1) \quad,$$

$$\text{(38d)} \quad G_{AB} \atop (3) = \text{diag}(1, 1, -1, -1, -1, 1) \quad.$$

The tensors (38) are not independent: there remains one degree of freedom in the choice of the coefficients $Z_{(i)}$ . One obtains easily

(39a) $\quad Z_{(1)} = (3\, Z_{(0)} + \nu_1 - \nu_2 - \nu_3 )\, /\, 2 \quad,$

(39b) $\quad Z_{(2)} = (3\, Z_{(0)} - \nu_1 + \nu_2 - \nu_3 )\, /\, 2 \quad,$

(39c) $\quad Z_{(3)} = (3\, Z_{(0)} - \nu_1 - \nu_2 + \nu_3 )\, /\, 2 \quad.$

If the medium is isotropic, we have

(40) $\quad \nu_1 = \nu_2 = \nu_3 = \nu \quad.$

Then, one can choose $Z_{(0)}$ in order to keep only the first term and obtain the classical expression (4), let's be

(41) $\quad Z_{(0)} = \nu\, /\, 3 \quad.$

Then, one has

(42) $\quad Z_{(1)} = Z_{(2)} = Z_{(3)} = 0$

and, thus, the decomposition becomes (4) when identifying $Z_{(0)}$ with $Z_0$ .

## Decomposition of the second diagonal of the constitutive tensor

We name "second diagonal" the diagonal in (33) formed by $\mu_{(1)}$ , $\mu_{(2)}$ and $\mu_{(3)}$ . That diagonal can also be decomposed. We define for that

(43)    $Z_{(4)} = (\mu_1 + \mu_2 + \mu_3 ) / 3$

and extract the corresponding term of $\chi_{AB}$ , namely

(44)    $Z_{(4)} \, \varepsilon_{AB}$  .

The term (44) takes into account the completely anti-symmetric part of $\chi_{\mu\nu\rho\sigma}$ . That term is the so-named "axion part" in the literature [5][7]. All the other terms of this paragraph and of the precedent paragraph are thus parts of the so-called "principal part" of the constitutive tensor [5][7].

There remain then only two degrees of freedom. For dealing with those, let's define

(45)    $g_{\mu\nu} \atop (5) = \mathrm{diag}(1, 1, 1, 1)$

We define also

(46a)    $g_{\mu\nu} \atop (6) = \begin{bmatrix} 0 & 1 & 0 & 0 \\ 1 & 0 & 0 & 0 \\ 0 & 0 & 0 & 1 \\ 0 & 0 & 1 & 0 \end{bmatrix}$ ,

(46b)    $g_{\mu\nu} \atop (7) = \begin{bmatrix} 0 & 1 & 0 & 0 \\ 1 & 0 & 0 & 0 \\ 0 & 0 & 0 & -1 \\ 0 & 0 & -1 & 0 \end{bmatrix}$ ,

(46c)    $g_{\mu\nu} \atop (8) = \mathrm{diag}(-1, -1, 1, 1)$    ,

(47a)    $g_{\mu\nu} \atop (9) = \begin{bmatrix} 0 & 0 & 1 & 0 \\ 0 & 0 & 0 & 1 \\ 1 & 0 & 0 & 0 \\ 0 & 1 & 0 & 0 \end{bmatrix}$ ,

(47b) $\quad g_{\mu\nu} \atop (10) = \begin{bmatrix} 0 & 0 & 1 & 0 \\ 0 & 0 & 0 & -1 \\ 1 & 0 & 0 & 0 \\ 0 & -1 & 0 & 0 \end{bmatrix}$ ,

(47c) $\quad g_{\mu\nu} \atop (11) = \mathrm{diag}(-1, 1, -1, 1)$ ,

(48a) $\quad g_{\mu\nu} \atop (12) = \begin{bmatrix} 0 & 0 & 0 & 1 \\ 0 & 0 & 1 & 0 \\ 0 & 1 & 0 & 0 \\ 1 & 0 & 0 & 0 \end{bmatrix}$ ,

(48b) $\quad g_{\mu\nu} \atop (13) = \begin{bmatrix} 0 & 0 & 0 & 1 \\ 0 & 0 & -1 & 0 \\ 0 & -1 & 0 & 0 \\ 1 & 0 & 0 & 0 \end{bmatrix}$ ,

(48c) $\quad g_{\mu\nu} \atop (14) = \mathrm{diag}(-1, 1, 1, -1)$ .

Several combinations of the ten $G_{AB} \atop (i)$ associated with (45) to (48) have non-zero elements only on the second diagonal.

For example, the terms

(48d) $\quad G_{AB} \atop (6) = \begin{bmatrix} -1 & 0 & 0 & 0 & 0 & 0 \\ 0 & 0 & 0 & 0 & -1 & 0 \\ 0 & 0 & 0 & 0 & 0 & 1 \\ 0 & 0 & 0 & -1 & 0 & 0 \\ 0 & -1 & 0 & 0 & 0 & 0 \\ 0 & 0 & 1 & 0 & 0 & 0 \end{bmatrix}$ ,

(48 e) $\quad G_{AB} \atop (7) = \begin{bmatrix} -1 & 0 & 0 & 0 & 0 & 0 \\ 0 & 0 & 0 & 0 & 1 & 0 \\ 0 & 0 & 0 & 0 & 0 & -1 \\ 0 & 0 & 0 & -1 & 0 & 0 \\ 0 & 1 & 0 & 0 & 0 & 0 \\ 0 & 0 & -1 & 0 & 0 & 0 \end{bmatrix}$ ,

are not in a suitable form, in contrast to

$$(49) \quad \underset{(6)}{G_{AB}} - \underset{(7)}{G_{AB}} = \begin{bmatrix} 0 & 0 & 0 & 0 & 0 & 0 \\ 0 & 0 & 0 & 0 & -2 & 0 \\ 0 & 0 & 0 & 0 & 0 & 2 \\ 0 & 0 & 0 & 0 & 0 & 0 \\ 0 & -2 & 0 & 0 & 0 & 0 \\ 0 & 0 & -2 & 0 & 0 & 0 \end{bmatrix}.$$

One obtains also other expressions of a suitable form, such as

$$(50) \quad \underset{(9)}{G_{AB}} - \underset{(10)}{G_{AB}} \quad ,$$

$$(51) \quad \underset{(12)}{G_{AB}} - \underset{(13)}{G_{AB}} \quad ,$$

$$(52) \quad \underset{(5)}{G_{AB}} + \underset{(6)}{G_{AB}} + \underset{(9)}{G_{AB}} + \underset{(13)}{G_{AB}} \quad ,$$

$$(53) \quad \underset{(8)}{G_{AB}} + \underset{(6)}{G_{AB}} - \underset{(9)}{G_{AB}} - \underset{(13)}{G_{AB}}$$

and other similar sums.

It is then easy to achieve the decomposition of $\chi_{AB}$ in terms of the requested form. However, the decomposition is again not unique.

## Conclusions

We have provided a method in order to obtain a decomposition of the form (33) required in [8], for the electromagnetic constitutive tensor and thus for the associated area metric. We also obtain decomposition in the form required in [8]. That decomposition is a deeper splitting of the so-called principal part of the electromagnetic constitutive tensor [5][7]. Unfortunately, the decomposition is not unique. Consequently, the method proposed in [8] probably provides not one affine connection, but a class of such connection.

However, the tool provided in this paper is powerful because it allows the decomposition of the electromagnetic constitutive tensor without reference to any pre-existent metric.

## Appendix I

The quantities $F_{\mu\nu}$ and $H^{\mu\nu}$ have a simple interpretation after time-space splitting. Let's assume that the first co-ordinate is in time direction and that the three other co-ordinates are spatially oriented. Then, the components of these quantities are

$$(54) \quad F_{\mu\nu} = \begin{bmatrix} 0 & -E_1 & -E_2 & -E_3 \\ E_1 & 0 & B^3 & -B^2 \\ E_2 & -B^3 & 0 & B^1 \\ E_3 & B^2 & -B^1 & 0 \end{bmatrix}$$

and

$$(55) \quad H^{\mu\nu} = \begin{bmatrix} 0 & D^1 & D^2 & D^3 \\ -D^1 & 0 & H_3 & -H_2 \\ -D^2 & -H_3 & 0 & H_1 \\ -D^3 & H_2 & -H_1 & 0 \end{bmatrix},$$

where

$E_1$, $E_2$ and $E_3$ are the components of the electric field strength,

$D^1$, $D^2$ and $D^3$ are the components of the electric displacement,

$H_1$, $H_2$ and $H_3$ are the components of the magnetic field (magnetising field),

$B^1$, $B^2$ and $B^3$ are the components of the magnetic flux density.

The fields E, D, H and B follow a formalism deduced from the Maxwell-Minkowski-Post formalism. It has proved to be useful in engineering [3][6].

## Appendix II

If we consider a frame (t, x, y, z) where the components of the metric are

$$(56) \quad g_{\mu\nu} = \begin{bmatrix} -c^2 & & & \\ & 1 & & \\ & & 1 & \\ & & & 1 \end{bmatrix} \quad \text{with} \quad c \quad \text{the light velocity,}$$

then, the constitutive tensor of the classical vacuum has the components (in the 6-vector convention)

$$(58) \quad \chi_{AB} = \begin{bmatrix} -\dfrac{1}{\varepsilon_o} & 0 & 0 & 0 & 0 & 0 \\ 0 & -\dfrac{1}{\varepsilon_o} & 0 & 0 & 0 & 0 \\ 0 & 0 & -\dfrac{1}{\varepsilon_o} & 0 & 0 & 0 \\ 0 & 0 & 0 & \mu_o & 0 & 0 \\ 0 & 0 & 0 & 0 & \mu_o & 0 \\ 0 & 0 & 0 & 0 & 0 & \mu_o \end{bmatrix},$$

where $\varepsilon_o$ and $\mu_o$ are respectively the electric permittivity and the magnetic permeability of the vacuum.